\documentstyle[twocolumn,prl,aps,psfig]{revtex}

\begin{document}
\title{Cooling of a mirror by radiation pressure}
\author{P.F.\ Cohadon, A.\ Heidmann, M.\ Pinard\thanks{%
e-mail : cohadon, heidmann or pinard@spectro.jussieu.fr}}
\address{Laboratoire Kastler Brossel\thanks{%
Laboratoire de l'Universit\'{e} Pierre et Marie Curie et de l'Ecole Normale
Sup\'{e}rieure associ\'{e} au Centre National de la Recherche Scientifique},
Case 74, 4 place Jussieu, F75252\ Paris Cedex 05, France}
\date{June 1999}
\maketitle

\begin{abstract}
We describe an experiment in which a mirror is cooled by the radiation
pressure of light. A high-finesse optical cavity with a mirror coated on a
mechanical resonator is used as an optomechanical sensor of the Brownian
motion of the mirror.\ A feedback mechanism controls this motion via the
radiation pressure of a laser beam reflected on the mirror.\ We have
observed either a cooling or a heating of the mirror, depending on the gain
of the feedback loop.\bigskip
\end{abstract}

\bigskip

Thermal noise is a basic limit for many very sensitive optical measurements
such as interferometric gravitational-wave detection\cite
{Saulson90,Bradaschia90,Abramovici92}. Brownian motion of suspended mirrors
can be decomposed into suspension and internal thermal noises.\ The latter
is due to thermally induced deformations of the mirror surface and
constitutes the major limitation of gravitational-wave detectors in the
intermediate frequency domain\cite{Bondu95,Gillespie95}. Observation and
control of this noise have thus become an important issue in precision
measurements\cite{Braginsky93,Bernardini98,Rugar91}. In order to reduce
thermal noise effects, it is not always possible to lower the temperature
and other techniques have been proposed such as feedback control\cite
{Mancini98}.

In this letter we report the first experimental observation of the cooling
of a mirror by feedback control. The principle of the experiment is to
detect the Brownian motion of the mirror with an optomechanical sensor and
then to freeze the motion by applying an electronically controlled radiation
pressure on the mirror. Mechanical effects of light on macroscopic objects
have already been observed, such as the dissipative effects of
electromagnetic radiation\cite{Braginsky70}, the optical bistability and
mirror confinement in a cavity induced by radiation pressure\cite{Dorsel83},
or the regulation of the mechanical response of a microcantilever by
feedback via the photothermal force\cite{Mertz93}. In our experiment the
radiation pressure is driven by the feedback loop in such a way that a
viscous force is applied to the mirror.\ It thus plays a role somewhat
similar to the one in optical molasses for atoms.

\begin{figure}
\centerline{\psfig{figure=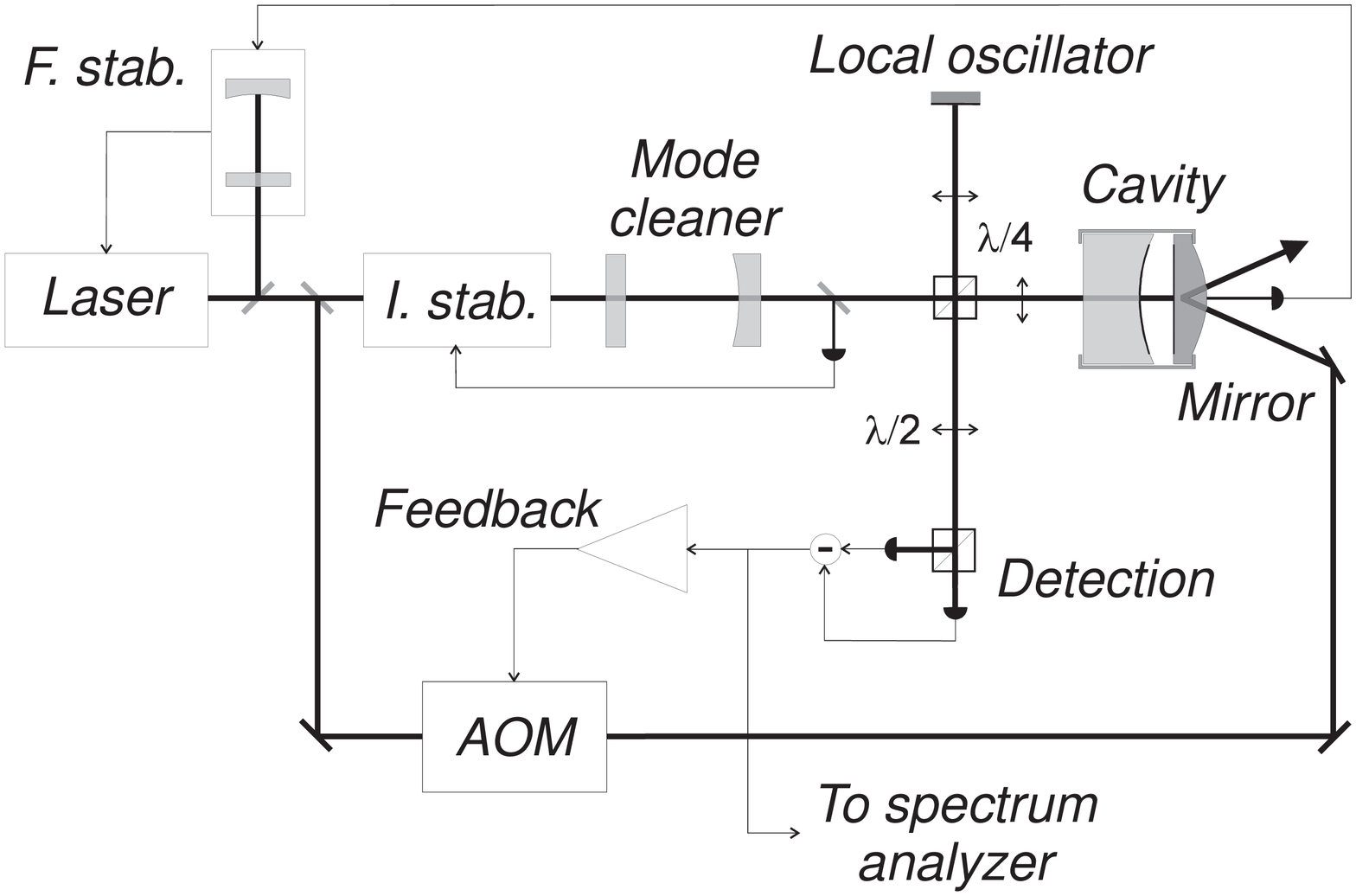,width=7.5cm}}
\caption{Experimental setup. The Brownian motion of the mirror is detected with a
high-finesse cavity. A frequency (F. stab.) and intensity (I. stab.) stabilized light beam
is sent into the cavity and the phase of the reflected field is measured by homodyne
detection. This signal is fed back to the mirror via the radiation pressure exerted by
a beam with modulated intensity (AOM)}
\label{Fig_Setup}
\end{figure}

The cooling mechanism can be understood from the experimental setup shown in
figure \ref{Fig_Setup}. The mirror is used as the rear mirror of a
single-ended Fabry-Perot cavity.\ The phase of the field reflected by the
cavity is very sensitive to changes in the cavity length\cite
{Mio95,Tittonen99,Hadjar99}.\ For a resonant cavity, a displacement $\delta x
$ of the rear mirror induces a phase shift $\delta \varphi _{x}$ of the
reflected field on the order of 
\begin{equation}
\delta \varphi _{x}\approx \frac{8{\cal F}}{\lambda }\delta x\text{,}
\label{Eq_dpout}
\end{equation}
where ${\cal F}$ is the cavity finesse and $\lambda $ is the optical
wavelength. This signal is superimposed to the quantum phase noise of the
reflected beam.\ Provided that the cavity finesse is high enough, this
quantum noise is negligible and the Brownian motion of the mirror can be
detected by measuring the phase of the reflected field\cite{Hadjar99}.

To cool the mirror we use an auxiliary laser beam reflected from the back on
the mirror.\ This beam is intensity-modulated by an acousto-optic modulator
driven by the feedback loop so that a modulated radiation pressure is
applied to the mirror. The resulting motion can be described by its Fourier
transform $\delta x\left[ \Omega \right] $ at frequency $\Omega $ which is
proportional to the applied forces 
\begin{equation}
\delta x\left[ \Omega \right] =\chi \left[ \Omega \right] \left( F_{T}\left[
\Omega \right] +F_{rad}\left[ \Omega \right] \right) \text{,}  \label{Eq_dx}
\end{equation}
where $\chi \left[ \Omega \right] $ is the mechanical susceptibility of the
mirror. If we assume that the mechanical response is harmonic, this
susceptibility has a lorentzian shape 
\begin{equation}
\chi \left[ \Omega \right] =\frac{1}{M\left( \Omega _{M}^{2}-\Omega
^{2}-i\Gamma \Omega \right) }\text{,}  \label{Eq_Chi}
\end{equation}
characterized by a mass $M$, a resonance frequency $\Omega _{M}$ and a
damping $\Gamma $ related to the quality factor $Q$ of the mechanical
resonance by $\Gamma =\Omega _{M}/Q$.

The force $F_{T}$ in eq. (\ref{Eq_dx}) is a Langevin force responsible for
the Brownian motion of the mirror.\ At thermal equilibrium its spectrum $%
S_{F_{T}}\left[ \Omega \right] $ is related to the mechanical susceptibility
by the fluctuation-dissipation theorem 
\begin{equation}
S_{F_{T}}\left[ \Omega \right] =-\frac{2k_{B}T}{\Omega }%
\mathop{\rm Im}%
\left( \frac{1}{\chi \left[ \Omega \right] }\right) =2M\Gamma k_{B}T\text{,}
\label{Eq_ST}
\end{equation}
where $T$ is the temperature. The resulting thermal noise spectrum $S_{x}^{T}%
\left[ \Omega \right] $ of the mirror motion has a lorentzian shape centered
at frequency $\Omega _{M}$ and of width $\Gamma $.

The second force $F_{rad}$ in eq. (\ref{Eq_dx}) is the radiation pressure
exerted by the auxiliary laser beam and modulated by the feedback loop.
Neglecting the quantum phase noise in the control signal, this force is
proportional to the displacement $\delta x$ of the mirror (eq.\ \ref
{Eq_dpout}).\ We choose the feedback gain in such a way that the radiation
pressure is proportional to the speed $v=i\Omega \delta x$ of the mirror 
\begin{equation}
F_{rad}\left[ \Omega \right] =iM\Omega g\delta x\left[ \Omega \right] \text{,%
}  \label{Eq_Frad}
\end{equation}
where $g$ is related to the electronic gain. The radiation pressure exerted
by the auxiliary laser beam thus corresponds to an additional viscous force
for the mirror. The resulting motion is given by 
\begin{equation}
\delta x\left[ \Omega \right] =\frac{1}{M\left[ \Omega _{M}^{2}-\Omega
^{2}-i\left( \Gamma +g\right) \Omega \right] }F_{T}\left[ \Omega \right] 
\text{.}  \label{Eq_dxResult}
\end{equation}
This equation is similar to the one obtained without feedback (eq. \ref
{Eq_dx} with $F_{rad}=0$) except that the radiation pressure changes the
damping without adding any fluctuations. The noise spectrum $S_{x}^{fb}\left[
\Omega \right] $ of the mirror motion still has a lorentzian shape but with
a different width $\Gamma _{fb}=\Gamma +g$ and a different height.\ The
variation of height can be characterized by the amplitude noise reduction $%
{\cal R}$ at resonance frequency 
\begin{equation}
{\cal R}=\sqrt{\frac{S_{x}^{T}\left[ \Omega _{M}\right] }{S_{x}^{fb}\left[
\Omega _{M}\right] }}=\frac{\Gamma _{fb}}{\Gamma }=\frac{\Gamma +g}{\Gamma }%
\text{.}  \label{Eq_R}
\end{equation}
The resulting motion is then equivalent to a thermal equilibrium at a
different temperature $T_{fb}$ which can be either reduced or increased
depending on the sign of the gain $g$%
\begin{equation}
\frac{T_{fb}}{T}=\frac{\Gamma }{\Gamma _{fb}}=\frac{\Gamma }{\Gamma +g}\text{%
.}  \label{Eq_Teff}
\end{equation}

We now describe our experiment. The mirror is coated on the plane side of a
small plano-convex mechanical resonator made of silica (figure \ref
{Fig_Setup}).\ The coating has been made at the {\it Institut de Physique
Nucl\'{e}aire de Lyon} on a 1.5-mm thick substrate with a diameter of 14 mm
and a curvature radius of the convex side of 100 mm. Internal acoustic modes
correspond to gaussian modes confined around the central axis of the
resonator\cite{Wilson74,Pinard99}. The fundamental mode studied in this
paper is a compression mode with a waist equal to 3.4 mm and a resonance
frequency close to 2\ MHz\cite{Hadjar99,Pinard99}. The front mirror of the
cavity is a {\it Newport high-finesse SuperMirror} (curvature radius = 1 m,
transmission = 50\ ppm) held at 1\ mm from the back mirror. We have measured
the following parameters of the cavity : free spectral range = 141\ GHz,
cavity bandwidth = 1.9\ MHz, beam waist = 90 $\mu $m.\ These values
correspond to a finesse ${\cal F}$ of 37000.

The light entering the cavity is provided by a titane-sapphire laser working
at 810 nm and frequency-locked to a stable external cavity which is locked
to a resonance of the high-finesse cavity by monitoring the residual light
transmitted by the rear mirror.\ The beam is intensity-stabilized and
spatially filtered by a mode cleaner.\ One gets a 100-$\mu $W incident beam
on the high-finesse cavity with a mode matching of 98\%. The phase of the
field reflected by the cavity is measured by homodyne detection.\ The
reflected field is mixed with a 10-mW local oscillator and a servoloop
monitors the length of the local oscillator arm so that we measure the phase
fluctuations of the reflected field.\ This signal is
sent both to the feedback loop and to a spectrum analyzer.

The feedback loop consists of an amplifier with variable gain and phase
which drives the acousto-optic modulator. The 500-mW auxiliary beam is
uncoupled from the high-finesse cavity by a frequency shift of the
acousto-optic modulator (200\ MHz) and by a tilt angle of 10$%
{{}^\circ}%
$ with respect to the cavity axis. We have checked that this beam has no
spurious effect on the homodyne detection. A band-pass filter centered at
the fundamental resonance frequency of the mirror is also inserted in the
feedback loop to reduce its saturation. For large gains, the radiation
pressure $F_{rad}$ can become of the same order as the Langevin force $F_{T}$
and it must be restricted in frequency in order to get a finite variance.\ The
electronic filter has a quality factor of 200 and limits the efficiency of
the feedback loop to a bandwidth of 9 kHz around the fundamental resonance
frequency.

\begin{figure}
\centerline{\psfig{figure=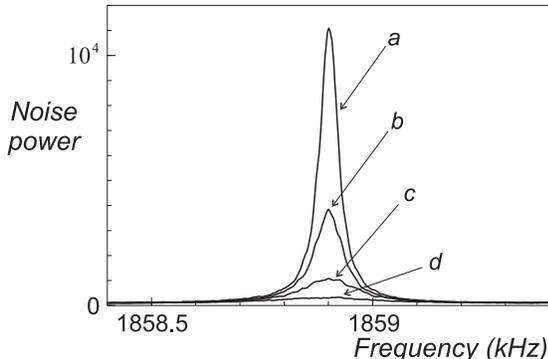,height=4.8cm}}
\caption{Phase noise spectrum of the reflected field normalized to the shot-noise level
for a frequency span of 1 kHz around the fundamental resonance frequency of the
mirror.\ The peak reflects the Brownian motion of the mirror without feedback
({\it a}) and with feedback for increasing gains ({\it b} to {\it d})}
\label{Fig_Cooling}
\end{figure}

Figure \ref{Fig_Cooling} shows the phase noise spectrum of the reflected
field obtained by an average of 1000 scans of the spectrum analyzer with a
resolution bandwidth of 10\ Hz.\ Curve (a) is obtained at room temperature
without feedback. It reproduces the thermal noise spectrum $S_{x}^{T}\left[
\Omega \right] $ of the mirror which is concentrated around the fundamental
resonance frequency (1858.9 kHz) with a width $\Gamma /2\pi $ of 45 Hz
(mechanical quality factor $Q\simeq 40000$). The spectrum is normalized to
the shot-noise level and it clearly appears that the thermal noise is much
larger than the quantum phase noise\cite{Hadjar99}.

Curves (b) to (d) are obtained with feedback for increasing electronic
gains. The phase of the amplifier is adjusted to maximize the correction at
resonance.\ From eqs. (\ref{Eq_Frad}) and (\ref{Eq_dxResult}) this
corresponds to a global imaginary gain for the loop and to a purely viscous
radiation pressure force. The control of the mirror motion is clearly
visible on those curves. The thermal peak is strongly reduced while its
width is increased. The amplitude noise reduction ${\cal R}$ at resonance is
larger than 20 for large gains.

The effective temperature $T_{fb}$ can be deduced from the variance $\Delta
x^{2}$ of the mirror motion which is equal to the integral of the spectrum $%
S_{x}^{fb}\left[ \Omega \right] $.\ From eqs. (\ref{Eq_ST}), (\ref
{Eq_dxResult}) and (\ref{Eq_Teff}) one gets the usual relation for a
harmonic oscillator at thermal equilibrium 
\begin{equation}
\frac{1}{2}M\Omega _{M}^{2}\Delta x^{2}=\frac{1}{2}k_{B}T_{fb}\text{.}
\label{Eq_Variance}
\end{equation}
The decrease of the area of the thermal peak observed in figure \ref
{Fig_Cooling} thus corresponds to a cooling of the mirror.

\begin{figure}
\centerline{\psfig{figure=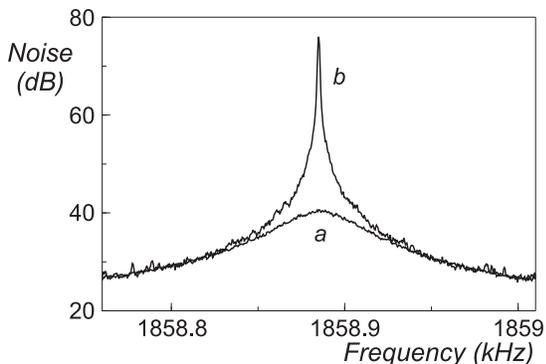,height=4.8cm}}
\caption{Phase noise spectrum of the reflected field in dB scale normalized to the
shot-noise level for a frequency span of 250 Hz, without feedback ({\it a}) and with
feedback for a negative gain ({\it b})}
\label{Fig_Heating}
\end{figure}

Figure \ref{Fig_Heating} shows the effect of feedback for a reverse gain ($%
g<0$). Noise spectra are obtained by an average of 500 scans with a
resolution bandwidth of 1\ Hz.\ Curve (b) exhibits a strong increase of the
thermal peak which now corresponds to a heating of the mirror.\ The feedback
also reduces the damping from $\Gamma $ to $\Gamma -\left| g\right| $, thus
increasing the quality factor of the resonance.\ We have obtained a maximum
effective quality factor of $2.2\times 10^{6}$ ($\Gamma _{fb}\simeq \Gamma
/50$), limited by the saturation of the feedback loop.

\begin{figure}
\centerline{\psfig{figure=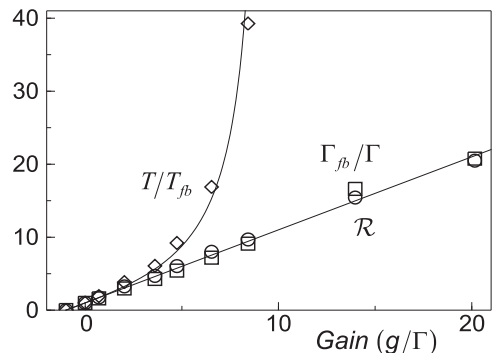,width=6.4cm}}
\caption{Variation of the damping $\Gamma _{fb}/\Gamma $ (squares), of the
amplitude noise reduction ${\cal R}$ at resonance (circles) and of the cooling
factor $T/T_{fb}$ (diamonds), as a function of the feedback gain $g$ normalized to
$\Gamma $.\ Solid curves are theoretical results}
\label{Fig_Gain}
\end{figure}

It is instructive to study the efficiency of the cooling or heating
mechanism with respect to the gain of the feedback loop. Figure \ref
{Fig_Gain} shows the variation of the damping $\Gamma _{fb}/\Gamma $, of the
amplitude noise reduction ${\cal R}$ at resonance and of the cooling factor $%
T/T_{fb}$, as a function of the feedback gain.\ These parameters are derived
from the experimental spectra by lorentzian fits which give the width and
the area of the thermal peak, the latter being related to the effective
temperature by eq. (\ref{Eq_Variance}). To measure the feedback gain, we
detect the intensity of the auxiliary beam after reflection on the mirror.\
The ratio between the modulation spectrum of this intensity at frequency $%
\Omega _{M}$ and the noise spectrum $S_{x}^{fb}\left[ \Omega _{M}\right] $
is proportional to the gain $g$.\ This measurement takes into account any
nonlinearity of the gain due to a possible saturation of the acousto-optic
modulator.

As expected from eqs. (\ref{Eq_dxResult}) and (\ref{Eq_R}), the damping and
the amplitude noise reduction ${\cal R}$ have a linear dependence with the
gain, as well for cooling ($g>0$) as for heating ($g<0$).\ The straight line
in figure \ref{Fig_Gain} is in excellent agreement with experimental data
and allows to normalize the gain $g$ to the damping $\Gamma $, as this has
been done in the figure.

This figure also shows that large cooling factors $T/T_{fb}$ can be
obtained. This cooling factor does not however evolve linearly with the gain
as it would be expected from eq. (\ref{Eq_Teff}). This is due to the
presence of a background thermal noise visible in figure \ref{Fig_Cooling}.\
This noise is related to all other acoustic modes of the mirror and to the
thermal noise of the coupling mirror of the cavity.\ The feedback loop has
not the same effect on this noise and on the fundamental thermal peak.\ The
solid curve in figure \ref{Fig_Gain} corresponds to a theoretical model in
which the background noise is assumed to be unchanged by the feedback.\ As a
consequence, only the fondamental mode is cooled at a temperature $T_{fb}$
whereas all other modes stay in thermal equilibrium at the initial
temperature $T$. The resulting cooling factor $T/T_{fb}$ is in excellent
agreement with experimental data.

\begin{figure}
\centerline{\psfig{figure=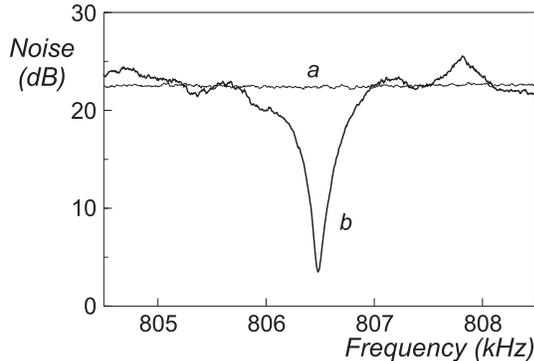,height=4.8cm}}
\caption{Cooling far from resonance (800 kHz). The background thermal noise ({\it a})
is reduced in presence of feedback ({\it b}). Curves represent the phase noise spectra
of the reflected field in dB scale normalized to the shot-noise level}
\label{Fig_CoolLF}
\end{figure}

The cooling mechanism is not limited to the mechanical resonance
frequencies.\ Figure \ref{Fig_CoolLF} shows the cooling obtained at
frequencies well below the fundamental resonance frequency. The electronic
filter is now centered around 800 kHz and the feedback loop reduces the
background thermal noise (curve a). The width of the noise reduction (curve
b) is related to the filter bandwidth. The phase and the gain of the
feedback loop has been adjusted since the electronic gain $g$ has now to be
compared to the real part $M\Omega _{M}^{2}$ of the inverse of the
mechanical susceptibility at low frequency (eq. \ref{Eq_dxResult}).\ Note
that the amplitude of the radiation pressure exerted by the auxiliary laser
beam is however approximately the same as in the resonant case.\ Large noise
reduction is actually obtained when the radiation pressure $F_{rad}$ is on
the order of the Langevin force $F_{T}$ whose amplitude is independent of
frequency (eq.\ \ref{Eq_dx}).

\begin{figure}
\centerline{\psfig{figure=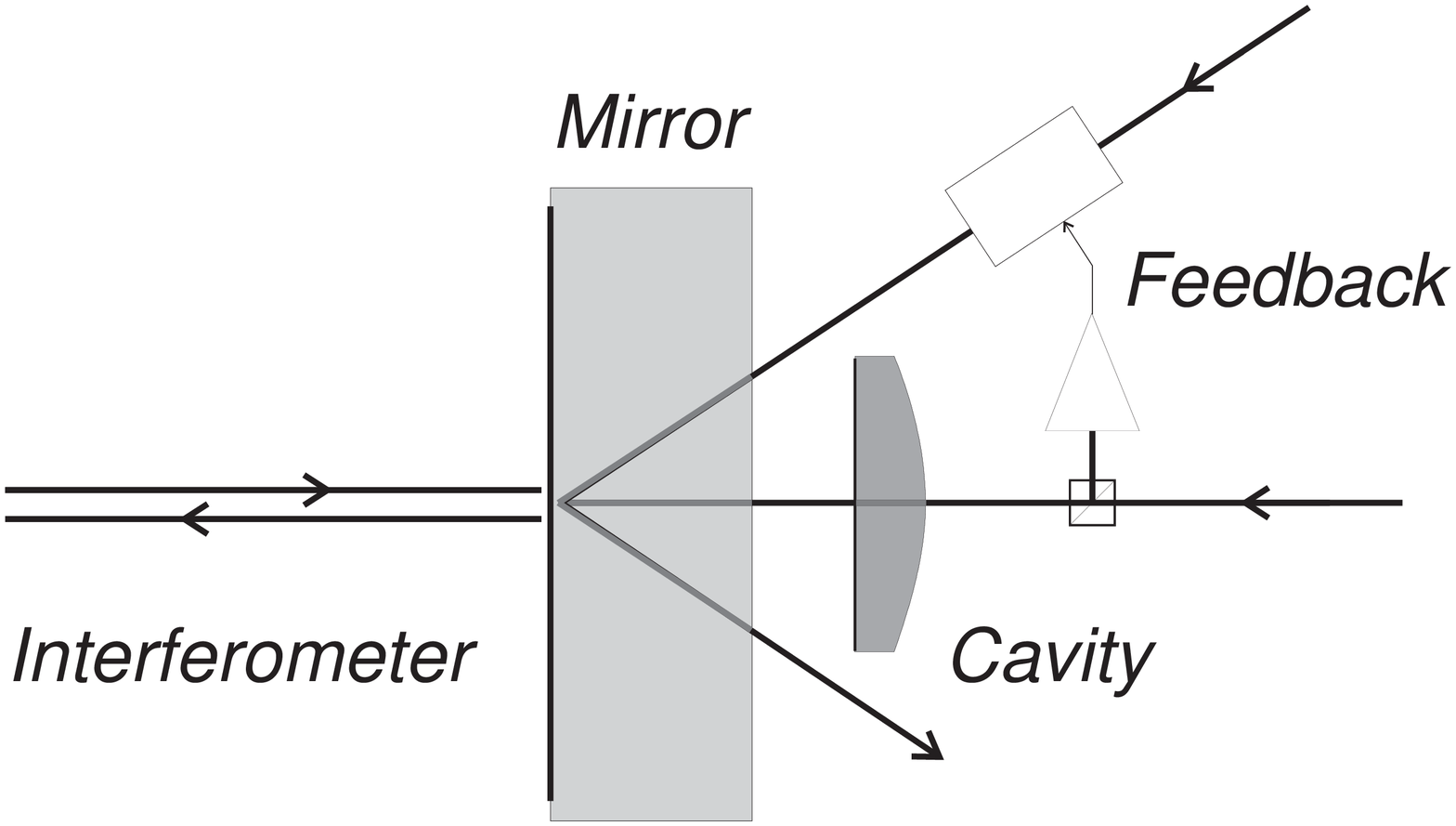,width=5.5cm}}
\caption{Application of the cooling mechanism to reduce the thermal noise in a
gravitational-wave interferometer}
\label{Fig_Control}
\end{figure}

In conclusion, we have observed a thermal noise reduction by a factor 20\
near the fundamental resonance frequency.\ The radiation pressure exerted by
the feedback loop corresponds to a viscous force which increases the damping
of the mirror without adding thermal fluctuations. For large gains, the
thermal peak of the fundamental mode becomes of the same order as the
background thermal noise and no global thermal equilibrium is reached. As
far as an effective temperature can be defined for the fundamental mode, we
have obtained a reduction of this temperature by a factor 40. We have also
observed a heating of the mirror for reverse feedback gains, and a cooling
of the background thermal noise at low frequencies.

The cooling mechanism demonstrated in this paper may be useful to increase
the sensitivity of gravitational-wave interferometers.\ The main difficulty
is to freeze the thermal noise without changing the effect of the signal.\
We propose in figure \ref{Fig_Control} a possible scheme to control the
thermal noise of one mirror of the interferometer. A cavity performs a local
measurement of the mirror motion which is fed back to the mirror via the
radiation pressure of a laser beam. The coupling mirror of the cavity is a
small plano-convex mirror with a high mechanical resonance frequency and a
low background thermal noise at low frequency. As a consequence the cavity
measures the thermal noise of the mirror of the interferometer.\ For a short
cavity, this measurement is not sensitive to a gravitational wave and the
cooling can reduce the background thermal noise at the gravitational-wave
frequencies without changing the response of the interferometer.

\end{document}